\documentclass[preprint, 12pt]{elsarticle}
\usepackage{amsmath}
\usepackage{graphicx}
\usepackage{epstopdf}

\pdfoutput=1 

\begin{document}

\begin{frontmatter}
\title{On the possible connection between inertia and intrinsic angular momenta and its consequences.}

\author{Giovanni Organtini}
\ead{giovanni.organtini@roma1.infn.it}
\address{ "Sapienza", Universit\`a di Roma \& INFN-Sez. di Roma, I-00185 ROMA, Italy}


\begin{abstract}
In this paper we show the consequences of a principle, according to which the dynamics of the Universe must not depend on the number of the particles of which it is composed. The validity of such a principle lead us to the conclusion that inertia and intrinsic angular momenta are deeply interrelated between them. In particular, assuming the principles outlined in the paper, matter must be composed by fermions, all stable bosons must be vectors and massless and no scalar particles can exist in the Universe. We also apply the results to the holographic principle and found established results more naturally with respect to previous approaches as well as new predictions about the Unruh effect.
\end{abstract}

\end{frontmatter}

\section{Introduction.}\label{introduction.}
Besides the known forces acting in the Universe, inertia is perhaps the most fundamental and, at the same time, the least understood one. Despite its very fundamental nature, inertial forces escape a reasonable explanation and are generically and qualitatively ascribed to the distribution of matter in the Universe. The well known Newton's bucket experiment illustrates very well this conviction. This simple, qualitative statement means that, in some unknown way, the water in the bucket is able to determine its position with respect to some external reference frame, defined by objects out of the bucket. Those objects missing, one cannot define any reference frame with respect to which the position of the molecules of the water can be measured. This statement can be regarded as one of the possible formulation of the Mach's principle and, in turn, means that there should be some sort of interaction between the water in the bucket and all the other bodies in the Universe, whatever the distance between those objects and the bucket. Without such interaction it was impossible for the water to {\em tell} its position with respect to any reference frame. It is in fact impossible even to define a reference frame. As a result, it is impossible, for the water, to climb the walls of the bucket. The water {\em does not know} wether it is rotating or not.

In this paper we discuss a very simple Universe, constituted by a small number of particles, and show that, in order for them to obey very basic principles, one must introduce some constraint about their intrinsic angular momentum. These observations have consequences on the nature of the particles and, in particular, forbid scalar particles as possible constituents of the Universe. 

Moreover, given the considerations above, we propose an alternative formulation of the principles outlined in~\cite{verlinde}, in which Erik Verlinde tries to introduce inertia as an entropic force. 

\section{Simple Universes.}
The simplest Universe we can imagine is composed by just one particle. The dynamics of such a Universe is trivial. No interaction is expected in it. There is no interest in studying a Universe like that. For interactions to appear, at least another particle is needed, irrespective of its nature. 

A more interesting Universe is the one composed by just two particles. Such a Universe, in fact, can be approximated by a group of two particles far enough from any other matter in the Universe. If there is no interaction between those two particles, the dynamics is still trivial. They just keep their state forever, i.e. they continue moving at constant speed along a linear path.

If there is some interaction there can be two different cases, depending on the relative state. The two particles can either approach each other, if the interaction is attractive, or their distance increases because the interaction is repulsive. If the relative state of the particles is such that they have some non--vanishing angular momentum, the trajectory of the particles will be a curve. Eventually they can orbit around a common center of mass.

However, seen from a reference frame attached to one of the particles, the interaction can only be described in terms of the distance between the two particles. In fact, since there are no other objects in the Universe, from the point of view of one particle, the only meaningful physics quantity that can be measured is the distance with respect to the other particle. In other words, such a Universe appears to be unidimensional. The only direction that can be defined at any time by each particle is the one corresponding to the vector $\mathbf{r}$ connecting the two particles.

As a consequence, both particles cannot {\em tell} that they are rotating or changing their direction. Each particle can only detect a reduction or an enlargement of the distance with respect to the other. The dynamics of the particles, then, appear as a variation of the distance between the two particles. However, for rotating particles this rather simple observation leads to a contradiction. For the sake of simplicity, consider a particle $A$ as being at rest, while particle $B$ rotates around $A$ along a circular trajectory centered in $A$\footnote{Any other motion can be decomposed into a motion along a path along the radius and a path along an arc of circumference.}. From the point of view of $B$, the distance with respect to $A$ is constant. 

If particle $B$ finds itself at constant distance from $A$, it must conclude that there is no interaction between the two particles. On the other end, if we assume that particle $A$ produces some sort of field, with which $B$ interacts, the latter must be able to {\em measure} the intensity of such a field and it cannot appear to be null, just because $B$ appears to be at rest. In fact, let's consider a particle $B$ moving from infinite distance toward $A$, with a given impact parameter $d > 0$. While approaching $A$, particle $B$ feels the force produced by the $A$ field, and changes its trajectory. Eventually it can be captured by $A$ and $B$ can start orbiting around $A$. In this case, from the point of view of $B$, the $A$ field must vanish at a certain time. Unless it can experiment, in its reference frame, an inertial centrifugal force, depending on its acceleration, which make the sum of the forces to vanish. But in order to experiment such a force, the particle must be able to {\em tell} that it is rotating around some axis, and this is impossible in this case. The {\em rotating} particle, in fact, can just move forth and back with respect to the particle at rest. One cannot even tell that this is in fact a rotating particle. 

We arrived at a paradox: for particle $B$ there must be some field, but it results in null forces. If we believe that the laws of the dynamics must be valid irrespective of the number of particles in the Universe, we are lead to conclude that either a Universe must be constituted by at least three particles, or that there must be some other mean, for a particle, to tell its relative state with respect to another particle.

\section{Angular Momentum.}
In fact, if the two particles in our simple Universe were not scalars, but had a non--vanishing, intrinsic angular momentum, the paradox would disappear. If at least one of the particles has a non--null, conserved angular momentum $\mathbf{J}$, then its relative state with respect to the other particle depends upon both the distance $\mathbf{r}$ and the relative orientation of $\mathbf{J}$ with respect to $\mathbf{r}$. Only in one case there is still an ambiguity: it is the case in which $\mathbf{J}$ is perpendicular to $\mathbf{r}$.

If particle $B$ has an intrinsic, conserved angular momentum $\mathbf{J}$, despite the distance $r=|\mathbf{r}|$ remains constant, $\mathbf{J}$ precedes around the angular velocity of the particle, so that there is a mean for particle $B$ to measure how and how fast it is moving with respect to particle $A$. In this case, inertial centrifugal forces may appear and can cancel field generated forces, keeping particle $B$ at rest in its reference frame.

It seems, then, that intrinsic angular momenta are essential in introducing inertia in a Universe. For inertial forces to appear, even in the simplest possible Universe, matter particles must have some intrinsic angular momentum. In fact, matter particles need to have some non--vanishing intrinsic angular momentum along any possible direction, to avoid the case for which $\mathbf{J}$ is perpendicular to  $\mathbf{r}$. 

Such a requirement is satisfied if matter particles are fermions. In fact, in this case, whatever the direction along which the angular momentum is measured, its value is always different from zero. This observation is of capital importance. It represent a justification of the experimental fact that matter appears always in the form of fermions. In other words, if the above arguments are true, matter particles have to be fermions because this is required for a consistent inertial behavior.

\section{Photons, and other bosons.}
If it is true that matter particles appear as fermions in the Universe, it is also true that interactions are modeled as the exchange of bosons for all known forces. The electromagnetic interactions is carried by photons, weak force is mediated by $Z$ and $W^\pm$ bosons, while gluons are responsible for the strong interaction. All the mediators are bosons.

In order for the above mentioned considerations to be valid in this picture, bosons must obey the same rules of fermions, i.e. they must show non--vanishing intrinsic angular momenta in any possible direction. However, for bosons is allowed to have their spin perpendicular to $\mathbf{r}$ in such a way that $J_z=0$ along some direction.

Rather than being a problem, such an observation corroborates our assumptions. In fact, gauge invariance is another experimental evidence, for electromagnetic interactions. Gauge invariance translates, in QED, in the fact that photons comes in only two possible spin states: $+1$ and $-1$. Transversely polarized photons, with $J_z=0$, are forbidden. That is not true for the massive intermediate vector bosons $Z$ and $W^\pm$. It must be noted, however, that all of those bosons are unstable. They decay after a very short time into a pair of fermions. No free, stable $Z$ or $W^\pm$ exist and possible, short enough, violations of the above mentioned principles are allowed within the Eisenberg indetermination principle. Gluons are considered to be massless and, as a result, they should also be longitudinally polarized, as photons. In fact, gluons do not exist as free particles, because of confinement. As a result, they in fact behave like weak interaction mediators for what concern their rotational properties.

\section{Scalar particles.}
The Standard Model of electroweak interactions, in its minimal formulation, predicts the existence of a scalar Higgs boson, to be responsible for the mass of the fermions as well as of the intermediate vector bosons. Extensions of the minimal Standard Model, still predict the existence of at least one scalar particle.

According to our picture, however, scalar particles are not allowed as components of the Universe. They can, in fact, appear as non--elementary particles. In our model, elementary, point--like particles, cannot exist in spinless state, because otherwise inertial laws cannot be applied to them. We can say that scalar bosons cannot {\em tell} that they are rotating. 

In fact, we can still admit the existence of scalar particles, at the price of admitting a different behavior for them. If we admit they exist, we are forced to consider them as a sort of new ether: either they are at rest or they move only along straight paths, implying difficulties in the case of a Universe of finite size. Of course, they can always exist as composites.

\section{Application to the Holographic Principle.}
According to some authors, space--time emerges as a holographic image of a Universe with more dimensions~\cite{ADD}. In particular, the author of~\cite{verlinde}, Erik Verlinde, assumes the holographic principle and derives the inertial force as an entropic force, emerging from the fact that the entropy of a system of two particles tend to increase if the two particles move with respect each other. Despite the fact that many of the conclusions in~\cite{verlinde} are debatable, this is a reasonable attempt to formally introduce inertia in the dynamics of two particles and could be a good starting point for further investigations in this field.

In that paper, an entropy $S$ for a particle of mass $m$ interacting with another particle at distance $r$ is introduced as

\begin{equation*}
S = Amr\,,
\end{equation*}
where $A$ is a constant. Inertial forces appear as entropic forces, due to an entropy gradient, as

\begin{equation*}
F = T\frac{\partial S}{\partial r}\,.
\end{equation*}
Here $T$ has the dimensions of a temperature defined to be as the temperature required to cause an acceleration $a$ on a particle of mass $m$, following the relationship known as the Unruh effect~\cite{unruh}, according to which, detectors accelerated in vacuum  will detect the presence of particles at temperature $T$ such that

\begin{equation}\label{eq:unruh}
T = \frac{1}{2\pi k_B}\frac{\hbar a}{c}\,,
\end{equation}
where $k_B$ is the Boltzmann constant, $c$ the speed of light and $\hbar$ the Planck constant. 
Given the relationships above, it is straightforward to show that inertia is an entropic force. In fact

\begin{equation*}
F = \frac{1}{2\pi k_B}\frac{\hbar a}{c} A m = ma\,,
\end{equation*}
provided that 

\begin{equation*}
A^{-1}= \frac{1}{2\pi k_B}\frac{\hbar}{c}\,.
\end{equation*}
A number of questions arise from the above model. First of all, the entropy $S$ is a scalar, but it must be constructed using a quantity that is naturally a vector: the distance $\mathbf{r}$ for which only the length is taken as an ingredient of $S$. Moreover, the inclusion of the constant $\hbar$ into a classical constant appears as unnatural, though it cancels in the end.

In this section we try to derive similar results, using our conclusion according to which particles must have a spin that is a multiple of 1/2. For definiteness, we consider two spin--1/2 particles orbiting around a common center of mass $O$, such that one of them is almost at rest in $O$, while the other follows a circular path around $O$.

The entropy $S$ must be a scalar built from both $\mathbf{r}$ and $\mathbf{J}$. The simplest combination who gives rise to a scalar is then the scalar product $\mathbf{r}\cdot \mathbf{J}$ between the two and we can define $S$ as

\begin{equation*}
S = \alpha m \mathbf{r}\cdot \mathbf{J} = \alpha m r J_r\,,
\end{equation*}
where $J_r$ is the component of $\mathbf{J}$ along $\mathbf{r}$. Being $\mathbf{J}$ conserved $S$ depends on the time $t$:

\begin{equation*}
S=\alpha m r \frac{\hbar}{2}\cos{(\omega t + \phi)}\,.
\end{equation*}
The work done by the force $F$ is given by

\begin{equation}\label{eq:force}
\mathbf{F}\cdot \mathbf{dr} = T dS=T \alpha m \frac{\hbar}{2}\cos{(\omega t + \phi)}dr.
\end{equation}
Imposing the second law of dynamics $F=ma$, we recover the result found in~\cite{verlinde}, provided that

\begin{equation*}
\alpha = \frac{2A}{\hbar}=\frac{4\pi k_B c}{\hbar^2}\,.
\end{equation*}
and $T$ is modified as

\begin{equation}\label{eq:unruh2}
T\to T' = \frac{1}{2\pi k_B}\frac{\hbar a}{c}\frac{1}{\cos{(\omega t+\phi)}}=\frac{T}{\cos{(\omega t+\phi)}}
\end{equation}
The latter equation can be interpreted as follows. Consider first the case $\omega=0$, i.e. linear acceleration only. For a particle accelerated along a path parallel to its intrinsic angular momentum, $\cos{\phi}=\pm 1$ and $T' \to \pm T$, recovering the Unruh result, apart from the sign, discussed below. For particles accelerated along a path perpendicular to its intrinsic momentum, $\cos{\phi}=0$ and $T' \to \infty$. Note that, thanks to the cancellation that occurs between numerator and denominator, the value for the force given by~(\ref{eq:force}) is not affected by the divergence in $T'$. The usual interpretation of the Unruh prediction is that accelerated detectors measure, in their reference frame, a field thermally distributed as a black body radiation at temperature $T$. Unruh detectors consist of any device with two states. In our case the detector is a point--like particle with spin 1/2. When $\cos{\phi}=0$, $\phi=\pi/2$ and the intrinsic angular momentum of the particle is perpendicular to the acceleration. The component of $\mathbf{J}$ along the acceleration, then, is null, that means that in fact the particle is in a superposition of states $J_z=+\frac{1}{2}$ and $J_z=-\frac{1}{2}$, with equal probabilities. The interaction with any external field of such a system provokes both absorption and stimulated emission, and being the two states equally populated, the net effect cannot be detected, since the final state will be the same of the initial state. In other words, we predict that no Unruh effect should manifest when a point--like polarized particle with spin is accelerated perpendicularly to its polarization vector. In fact there is a precise relationship between the temperature of the black body radiation seen by the accelerated point--like detector and the angle between the acceleration direction and the polarization vector of the detector, given by~(\ref{eq:unruh2}), that can be eventually experimentally verified.

The different signs of the temperature $T'$ reflect the fact that if the polarization vector is oriented along the acceleration, $T' >0$ and the detector experiences the absorption of the external Unruh field, being in one of its eigenstates. In the opposite case, the detector finds itself in the other possible eigenstate and experiences stimulated emission.

In the approach followed by Verlinde in his paper, conversely, $T$ as to be interpreted as the temperature associated to the holographic screen needed to cause the acceleration, and is inversely proportional to the amount $N$ of bits of information obtained by a particle close to an holographic screen. An infinite temperature corresponds, then, to null information. Again, this is reasonable as long as the information is exchanged in terms of a field able to cause transitions between states.

\section{Conclusion.}
In this paper we introduced a principle according to which the dynamics of any possible Universe does not depend on the number of particles contained in it. As a consequence of such a principle, we are forced to rule out the possibility to find elementary point--like scalar particles in the Universe. 

Having ruled out not only the possibility of finding scalar particles in a Universe, but even to find particles for which one component of their intrinsic angular momentum is null, turns out into the need, for matter particles, to behave like spinors, explaining the nature of matter fields. The same requirement forces stable bosons to be massless and, as a consequence, transversal.

Applying the principle to the holographic principle, moreover, we are able to introduce inertial forces as entropic forces, more naturally with respect to what is made in~\cite{verlinde}.  We assume the entropy $S$ to be a scalar built by the simplest combination of the only two meaningful vectors that can be defined in the simplest possible Universe. Moreover, the presence of the constant $\hbar$ in the definition of $T$ becomes natural, being the dynamics governed by the intrinsic angular momentum of the interacting particles.

Finally, we made a prediction about possible results of experiments aiming at detecting the Unruh radiation. The prediction states that the Unruh temperature must depend on the relative orientation between the acceleration and the intrinsic angular momentum of the detector and turns out to be reasonable in terms of the definition of the detector given by Unruh, as a device with two distinct quantum states.

We are convinced that the picture outlined above is still very naive and subject to many criticism, however, to use the terminology introduced by Imre Lakatos~\cite{lakatos}, it has a remarkable content of positive heuristic and, because of that, is susceptible of interesting improvements. In fact, just assuming very basic first principles, we are able to justify many different and well established experimental results, as well as to make predictions about new effects.

\section{Acknowledgements.}
I am indebted to my friend Dr. Donato Bini, at CNR and ICRA, who introduced me into the physics of gravitation and brought some effects predicted by General Relativity to my attention. I was lead to the thoughts given above, starting from these discussions. His continuous and constructive criticism made it possible, for me, to arrive at a coherent view of the matter presented in this paper.

\end{document}